\documentstyle[prl,aps,epsfig,tighten]{revtex}

\begin{document}

\twocolumn[
\hsize\textwidth\columnwidth\hsize\csname@twocolumnfalse\endcsname
\preprint{Submitted to PRL}
\title{ Dynamics of the Local Moment Induced by Nonmagnetic Defects in Cuprates }
\author{ W.A. MacFarlane, J. Bobroff, H. Alloul, P. Mendels,
 N. Blanchard } 
\address{ LPS, B\^{a}t. 510, Universit\'e Paris-Sud, UMR8502 CNRS, 91405, Orsay Cedex, France}
\author{G. Collin}
\address{LLB, CEN Saclay, CEA-CNRS, 91191 Gif-sur-Yvette, France}
\author{J.-F. Marucco}
\address{LCNS, Universit\'e Paris-Sud, 91405, Orsay, France}
\date{\today}
\maketitle

\begin{abstract}
We present a study of the spin dynamics of magnetic
defects induced by Li substitution of the plane Cu in the
normal state of YBa$_2$Cu$_3$O$_{6+x}$. The fluctuations of the coupled Cu magnetic
moments in the vicinity of Li are probed by near-neighbour $^{89}$Y {\it and}
$^7$Li NMR spin lattice relaxation.
The data indicates that the magnetic perturbation fluctuates as a single entity with a
correlation time $\tau$ which scales with the local
static susceptibility. This behaviour is reminiscent of the low $T$ Kondo state
of magnetic impurities in conventional metals. Surprisingly it
extends well above the ``Kondo'' temperature for the underdoped pseudogapped case.
\end{abstract}
 \twocolumn \vskip.5pc ]
\narrowtext

\typeout{74.62.Dh Substitutions, impurities}
\typeout{76.60.-k NMR}
\typeout{74.25.Ha Superconductivity and Magnetism}

The substitution of the Cu$^{2+}$ of metallic doped CuO$_{2}$ planes by
spinless Zn$^{2+}$ has been shown in the normal state of YBa$_{2}$Cu$_{3}$O$%
_{6+x}$(YBCO$_{6+x}$) to induce a local magnetic moment nearby\cite
{mah1,mah2}. The generality of this novel impurity response is supported by
similar experiments on the {\it heterovalent} spinless substitutions: Al$^{3+}$ in La$%
_{2-x}$Sr$_{x}$CuO$_{4}$ (LSCO)\cite{ish} and Li$^{+}$ in YBCO$_{6+x}$\cite
{li1}. Such induced magnetism does not occur in a simple Fermi liquid, but
it does have parallels in other strongly correlated insulating spin systems 
\cite{quantimp} as well as nearly magnetic metals, e.g. Co or Ni in Pd\cite
{RS5}. In the cuprates, the induced magnetic defect is thought to consist of
a strongly perturbed region in the immediate vicinity of the spinless site
together with the associated intrinsic magnetic response of the CuO$_{2}$
plane at further distances. In the underdoped region, the nearby Cu
moments are sufficiently decoupled from
the strongly correlated CuO$_{2}$ band to make a Curie-like\cite{pm}
contribution to the static susceptibility (in contrast to the intrinsic 
$\chi (T)$). This model of the static local structure of the defect provides
a reasonable explanation of the Y satellite NMR spectra\cite{mah1,mah2} as
well as the broadening of the host NMR by long range spin polarization\cite
{hostnmr}. Recently, we have shown that $^{7}$Li NMR of Li substituted YBCO
provides a very sensitive local probe of the surrounding induced
magnetism\cite{li1}. Accurate measurements of the $^{7}$Li NMR shift $K$
revealed that the static susceptibility in the vicinity of the spinless
defect follows a Curie Weiss $(T+\Theta )^{-1}$ temperature dependence,
leading us to suggest that $\Theta $ might be associated with a Kondo-like
crossover temperature.

An important way to study the depth of this analogy is through the spin
dynamics of the induced moment. The correlation time $\tau$ of a local spin
coupled to a metal is determined by spin-flip scattering
with the conduction band electrons. In conventional systems, e.g. Cu:Fe, 
the Kondo effect results from the anomalous
behaviour of this scattering below a characteristic (Kondo) temperature $T_{K} \sim \Theta$,
corresponding to a transition from weak to strong coupling in the sense of
renormalization group theory\cite{wilson}.
For $T < T_K$, i.e. in the strong coupling regime,
both the local static magnetic susceptibility
$\chi_0$ and $\tau$ are determined by a single energy scale $T_K$.
As a result, there is a well-established universal proportionality
between them\cite{henri}. This scaling breaks down for $T > T_{K}$
(in the weak-coupling regime)\cite{gs}, where $\tau $ follows a Korringa
$T^{-1}$ temperature dependence.

In correlated electronic systems very few attempts to study the dynamics of
induced moments have been made so far.
Theoretical investigations are also currently rather limited for
the metallic cuprates\cite{thin2}
(though more advanced in insulators\cite{thin}).
The fluctuations of the magnetic defect can often be investigated successfully
through measurements of the spin lattice relaxation times $T_{1}$ of nearby nuclei\cite{henri}.
In YBCO such measurements on $^{89}$Y satellites
corresponding to the near neighbours of Zn were limited to
the underdoped state\cite{mah1,mah2}.
In LSCO the $^{27}$Al NMR itself\cite{ish} was used, but the
results were ambiguous because neither the defect concentration ($z$) or hole
doping ($n$) dependence were investigated\cite{ish,allcomm}.

In this paper, we carry out the first comprehensive study of the induced
local magnetic dynamics using the lithium $^{7}T_{1}^{-1}$.
We find predominantly single impurity 
behaviour at low Li concentration and extract the $T$ dependence of $\tau $ .
Our analysis of the data is
supported by measurements of the $^{89}T_{1}^{-1}$ of the yttrium NMR
satellites in the underdoped region.
The measured $\tau$ indicates that the induced moment is strongly
coupled to the conduction band.
Moreover, we find the universal low temperature relation between $\chi _{0}$ and $\tau $
expected for a Kondo effect. Surprisingly, this correlation is maintained
for all hole concentrations and all temperatures, even far above
$\Theta$, where the weak-coupling limit is expected.

The samples used for this study have nominal Li concentration $z$ (used
throughout) defined by YBa$_{2}$(Cu$_{1-z}$Li$_{z}$)$_{3}$O$_{x}$. However,
the per plane Li concentration in our samples was found to be $\sim 0.85z$%
\cite{li1}. Underdoping was accomplished by deoxidation and overdoping by Ca$%
^{2+}$/Y$^{3+}$ substitution. Except for the Ca codoped sample, the powders
were aligned in Stycast 1266 epoxy\cite{isotropic}. The spin lattice
relaxation rate $T_{1}^{-1}$ of the I=3/2 $^{7}$Li nucleus was measured, in
the same spectrometer as Ref. \cite{li1}, using a saturation recovery
sequence with (for the aligned powders) the magnetic field parallel to the 
{\it c}-axis. The resonance is sufficiently narrow that, using short ($\sim $%
1$\mu $s) RF pulses, the entire spectrum was irradiated, and the resulting
single exponential recovery yielded the rates shown in Fig. 1.

\begin{figure}[t]
\centerline{\epsfig{file=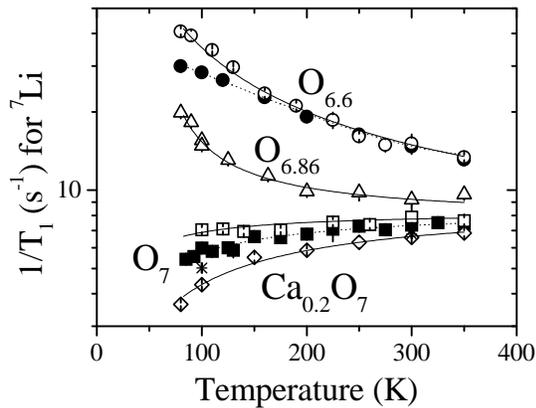,angle=0.0,width=7cm}}
\caption[]{
The spin lattice relaxation rate of Li substituted on Cu(2) in 
YBCO as a function of doping for
$z$=1\% (open symbols) and $z$=2\% (closed symbols) with
$\boldmath{H}\parallel c$.
The stars for $z$=2\% O$_7$ with $\boldmath{H}\perp c$
demonstrate the near isotropy of $T_1$. 
Note the weak $z$ dependence: a slightly higher rate
at low $T$ for smaller $z$.
}
\end{figure}
We can immediately make some important qualitative observations of the data: 
{\it i)} In our relatively dilute samples, we find no Li concentration
dependence of $^{7}T_{1}$ above 150K, and only a relatively small $z$
dependence below this. Thus for $z=1$\% the predominant nuclear relaxation
mechanism is characteristic of the isolated magnetic defect interacting with
the correlated metallic host. {\it ii)} The spin lattice relaxation rates of
the near neigbour $^{89}$Y NMR satellites\cite{li1} in the underdoped state
(YBCO$_{6.6}$) are quantitatively very similar to those of the Zn defect
(Fig. 2), indicating the generality of the dynamics of the induced magnetic
defect. {\it iii)} the $T$ and hole doping $n$ dependence of $T_{1}^{-1}$ is
unrelated to that of the NMR of the pure material, e.g. \cite{nmrrev}. Thus
we conclude that the dominant relaxation mechanism for Li is the
fluctuations of the magnetic defect and not the intrinsic magnetic
dynamics, as was also apparent for Zn\cite{mah1} and Al \cite{ish}.

\begin{figure}[h]
\centerline{\epsfig{file=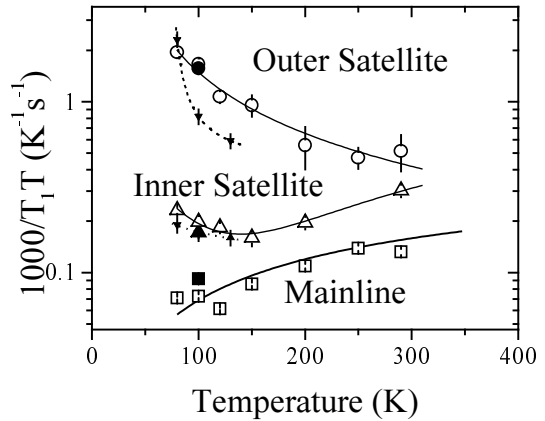,angle=0.0,width=7cm}}
\caption[]{
$1/T_1T$ for the satellite Y NMR of $z$=1\% (open points) and $z$=2\%
(large filled points) in YBCO$_{6.6}$.
Also shown (small filled points) are the results for the Y satellite
NMR of Zn doped samples from Ref. \cite{mah2}.
}
\end{figure}

To extract the timescale for fluctuations of the local moment, we employ a
relaxation model of the dynamic susceptibility, which is generally connected
to $1/T_1$ via the Moriya expression\cite{mor}: 
\begin{eqnarray}
\frac{1}{T_{1}} = \left( \frac{\mu _{n}}{\hbar 2\mu _{B}}\right) ^{2}k_BT {%
\sum_{{\bf q}} }\bar{A}^{2}({\bf q})\chi _{\perp}^{\prime \prime }({\bf q}%
,\omega _{n})/\omega _{n},
\end{eqnarray}
where $\mu_n$ is the nuclear magnetic moment, $\mu_B$ is the Bohr magneton, $%
\bar{A}$ is the (dominant) isotropic hyperfine coupling, $%
\chi^{\prime\prime}_{\perp}$ is the imaginary part of the perpendicular spin
susceptibility, and $\omega_n$ is the NMR frequency. In the superconducting state,
inelastic neutron scattering finds an impurity response peaked at the
antiferromagnetic wavevector with a characteristic energy of a few meV\cite{sidis}.
Thus at sufficiently low $T$, the defect will fluctuate as a single
magnetic entity\cite{allcomm} as there will be negligible occupation of
local internal fluctuation modes.
Assuming this limit, the expression above is
simply 
\begin{eqnarray}
\frac{1}{T_{1}} = 2 \left( \frac{\mu _{n}}{\hbar \mu _{B}}\right) ^{2}k_BT
A^{2} \chi _{\perp loc}^{\prime \prime }(\omega_{n})/\omega _{n},
\end{eqnarray}
where $A$ is the hyperfine coupling, with
\begin{eqnarray}
\chi _{\perp loc}(\omega )=\chi _{0}(T)\left\{ \frac{i-\omega \tau}
{i+(\omega -\omega _{e})\tau}\right\}.
\end{eqnarray}
Here $\tau$ is the correlation time, and $\omega_e$ is the EPR frequency. We
attribute the static macroscopic impurity susceptibility to the 4 near
neighbour sites per Li, i.e. $\chi_0=4\chi_{nn}$. The hyperfine coupling is
defined by its relation to the $^7$Li NMR shift, $K=A\chi_0/\mu_B$\cite{li1}.
The contribution to the macroscopic $\chi_0$ of more distant Cu sites, which
possess negligible hyperfine coupling to Li, make this a (probably slight)
overestimate of $A$. Assuming fast fluctuations, i.e. $\omega_e\tau \ll 1$,
yields 
\begin{eqnarray}
\frac{1}{T_{1}TK} = \left( \frac{2 k_B\mu _{n}^2 A}{\hbar^2 \mu _{B}}
\right) \tau.  \label{tauconv}
\end{eqnarray}

\begin{figure}[t]
\centerline{\epsfig{file=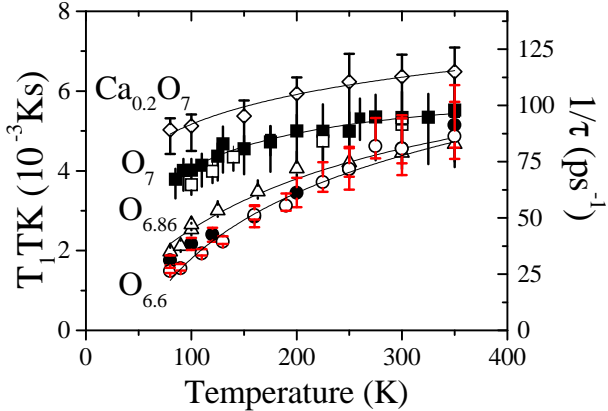,angle=0.0,width=8cm}}
\caption[]{
$T_1TK$ for Li in YBCO as a function of doping for
$z$=1\% (open symbols) and $z$=2\% (closed symbols).
$K$ is the magnetic shift. $T_1TK$ is
proportional to the inverse of the local moment correlation time $\tau$, right scale. 
}
\end{figure}
In Fig. 3 we plot $T_1TK$ vs. $T$ for all the doping levels, using $K$ from
Ref. \cite{li1}, and Eq.~(\ref{tauconv}) is used to make the right scale for 
$1/\tau$. The fluctuations measured by $\tau$ become slower as $n$ is
reduced towards the insulating magnetic phase. Moreover, the value is always
such that $\omega_e\tau \ll 1$ at the applied field of 7.5 T, as we
assumed\cite{epr}. 

Conventional treatments of Kondo impurities start from the exchange
hamiltonian $-J \vec{S} \cdot \vec{s}$ between the local spin $\vec{S}$ and
the band spins $\vec{s}$. In this case, the weak coupling limit
($T>T_K$) corresponds to a Korringa regime for
$\tau$, i.e.
$\hbar \tau^{-1} = \pi(J\rho)^2k_BT$, 
where $\rho$ is the density of states
at the Fermi level.
YBCO is not a simple metal, so
we do not expect simple Korringa behaviour. However,
$1/\tau$ for a weakly coupled magnetic moment should scale with the 
$1/T_1$ of the pure host nuclear spins, given the same hyperfine form factor.
This is the case, for example for Gd/Y substitution 
where the Gd ESR linewidth ($1/\tau_{{\it Gd}}$)
simply scales with the $1/T_1$ of yttrium \cite{jan}. It is clear that $1/\tau$
in Fig. 3 does not follow the Korringa Law, nor does
it in the $^{27}$Al experiment\cite{ish}. Moreover, in the underdoped case, 
where $T \gg \Theta$ the observed $1/ \tau$ displays none of the features of the
pseudogap detected by the $1/T_1$ of the planar oxygen or copper nuclei\cite{nmrrev}.
In contrast to
the Gd, this demonstates that the conventional weak-coupling regime is never evident.

On the other hand, because the induced moment is composed of
the same Cu atomic orbitals that normally participate in the band, we expect that it will
have a large overlap with band states and couple strongly to the band
excitations. In a conventional Kondo system,
all the properties of the low temperature Kondo state, corresponding to the
strong-coupling limit, are
determined by $T_K$; in particular, $\tau$ is proportional to
$\chi_0$. This limit ($T < \Theta$) can be attained in the optimal and
overdoped samples, where $\Theta$ exceeds 100 K.
Fig. 4 demonstrates such a correlation.
There appears to be some small deviation at
high $T$ (low shift), which we consider in more detail below.
Surprisingly, this scaling remains
valid for all $n$ {\it with the same slope}, even though $\Theta$ changes
drastically, i.e. from less than 5 K in O$_{6.6}$ to more than 200
K in the overdoped regime\cite{li1}. Thus, the strong coupling features of
the Kondo state persist even far above $\Theta$.

\begin{figure}[t]
\centerline{\epsfig{file=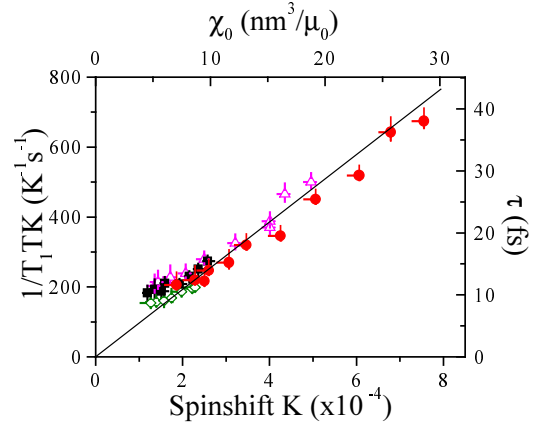,angle=0.0,width=7cm}}
\caption[]{
$1/T_1TK$ vs $K$ for $z$=1\% for various dopings: $%
x=6.6$ (full circles), $6.86$ (open triangles), $7$ (full squares) and $7$ Ca%
$_{0.2}$ (open diamonds). The proportionality of the static impurity
susceptibility and the correlation time of the moment is characteristic of
the $T < T_K$ Kondo regime. 
}
\end{figure}

Let us now consider the value of the 
proportionality constant between $\tau$ and $\chi_0$ which is determined independently of
the hyperfine coupling by 
\begin{eqnarray}
{\cal R} = {\frac{1}{T_{1}TK^2S} = \frac{\mu_e^2}{2\pi \hbar} \frac{\tau}{\chi_0}}
\label{korrconst}
\end{eqnarray}
where $S$ is the conventional Korringa constant, $4 \pi k_B / \hbar \times
(\mu_n / 2\mu_B)^2$. The values of ${\cal R}$ for Li are shown in Fig. 5. We
also include the corrected values for the Y satellites of Li (Fig. 2)\cite
{corr}.
The accuracy for the satellites is not as good as for the $^7$%
Li, but the values at low $T$ are consistent. The deviation at high $T$ in
Fig. 4 appears as the slight increase of ${\cal R}$ with $T$. This may be
the result of remnant impurity interactions yielding a slight decrease in $%
1/T_1$ from the isolated impurity value at low $T$, or it could signal the
apparition of another relaxation process at high $T$ possibly of quadupolar
origin. Assuming the defect consists of the 4 near neighbour sites, because
the Li is coupled to all 4 and each near-neighbour Y to only 2, if the 4
moments fluctuate independently $^{89}{\cal R}$ should be twice as large as $%
^{7}{\cal R}$, whereas they should be equal if the moments are rigidly
coupled. Despite the large error bars, the near equality at 100 K in Fig. 5
{\it favours the singly fluctuating model}. A relative increase of $^{89}{\cal R}$
over $^{7}{\cal R}$ at higher $T$ could also be due to internal magnetic excitations.

\begin{figure}[h]
\centerline{\epsfig{file=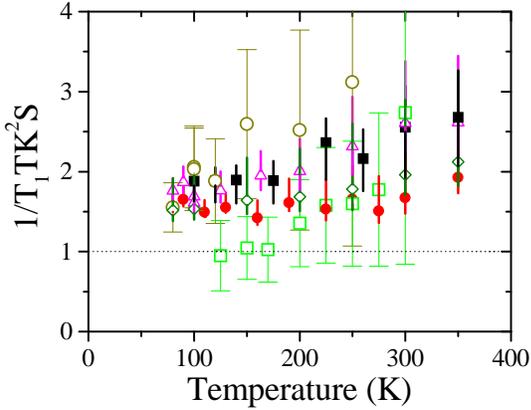,angle=0.0,width=7cm}}
\caption[]{
$1/T_1TK^2S$ vs. $T$
 for Li ($z$=1\%) for $x$=6.6 (full circles),
6.86 (open triangles), 7 (full squares) and 7 Ca$_{0.2}$ (open diamonds). Also
shown are the values for the outer Y satellite of the Li at $x$=6.6 (light open circles)
and for Al\cite{ish} (light open squares).
}
\end{figure}

For the induced moment in YBCO, ${\cal R}(T)$ is nearly constant,
independent of $n$, confirming that the defect doesn't change in character
from the underdoped far into the overdoped regime. We also include the data
from the Al experiment\cite{ish}, which possesses a remarkably similar $T$
dependence with a smaller value. 
In conventional Kondo systems, the crossover expected at $T_K$ should be reflected
in a marked decrease of ${\cal R}(t=T/T_K)$ from the zero temperature value through
$t = 1$ to a value determined by the parameter $(J\rho)^2$ for $t\gg 1$.
In the only case
where such measurements could be accomplished over a broad enough $t$ range
(Cu:Fe), the high $t$ value turned out to be only slightly less than the low 
$t$ limit\cite{henri}. In our case, $T_K$ varies strongly with $n$, but
${\cal R}(t)$ is nearly constant for $t\approx$ 0.4 to
100. This is further confirmation that the weak coupling limit is never attained. 

In summary, we have employed NMR as a powerful local probe of the dynamics
of the magnetic defect that is created by a Cu spin vacancy in the CuO$_2$
plane of YBCO. We deduced the correlation time for the induced magnetic
moment and concluded that it is strongly coupled to the band. As for a
conventional Kondo impurity, we find that the correlation time is simply
proportional to the static impurity susceptibility, making the case for
Kondo-like behaviour much more compelling than simply the behaviour of
$\chi_0(n,T)$\cite{li1}. However, in contrast to expectations from
conventional Kondo impurities, we find that this scaling extends far above 
$T_K$ and persists from the pseudogap to the overdoped regime.
Although the low $T$ behaviour mimics that of
the conventional Kondo state, $\Theta$ does not appear to be a crossover
from strong to weak coupling.
Such simple general behaviour of the defect response
suggests the possibility of a straightforward theoretical explanation, and
emphasizes the role of antiferromagnetic correlations
in the cuprate metallic state, even at high $T$ and far into the overdoped
regime.

WAM gratefully acknowledges the support of CIES, France and NSERC, Canada.
We have benefitted from discussions with Y. Sidis.



\begin{thebibliography}{99}
\bibitem{mah1}  A.V. Mahajan {\it et al.}, Phys. Rev. Lett. {\bf 72}, 3100
(94).

\bibitem{mah2}  A.V. Mahajan, H. Alloul, G. Collin and J.F. Marucco, Euro. Phys.
J. B, {\bf 13}, 457 (2000).

\bibitem{ish}  K. Ishida {\it et al.}, Phys. Rev. Lett. {\bf 76}, 531 (96).

\bibitem{li1}  J. Bobroff {\it et al.}, Phys. Rev. Lett. {\bf 83}, 4381 (99).

\bibitem{quantimp}  G.B. Martins {\it et al.}, Phys. Rev. Lett. {\bf 78},
3563 (97); 
K.M. Kojima {\it et al.}, Phys. Rev. Lett. {\bf 79}, 503 (97); 
F. Tedoldi, R. Santachiara and M. Horvatic, Phys. Rev. Lett. {\bf 83}, 412
(99); 
N. Fujiwara {\it et al.}, Phys. Rev. Lett. {\bf 80}, 604 (98).

\bibitem{RS5}  {\it Magnetism, Vol. V}, H. Suhl ed. (Academic Press, New
York, 1973) and references therein.

\bibitem{pm}  P. Mendels {\it et al.}, Europhys. Lett. {\bf 56}, 678 (99).

\bibitem{hostnmr}  H. Alloul {\it et al.}, Phys. Rev. Lett. {\bf 67}, 3140
(91); R.E. Walstedt {\it et al.}, Phys. Rev. B {\bf 48}, 10646 (93); J.
Bobroff {\it et al.}, Phys. Rev. Lett. {\bf 79}, 2117 (97).

\bibitem{wilson}  e.g. K.G. Wilson, Rev. Mod. Phys. {\bf 47}, 773 (75).

\bibitem{henri}  H. Alloul, Phys. Rev. Lett. {\bf 35}, 460 (75); J. Phys.
(Paris) {\bf 37}, L205 (76).

\bibitem{gs}  H. Shiba, Prog. Theor. Phys. {\bf 54}, 967 (75); 
W. G\"{o}tze and P. Schlottmann, Sol. St. Comm. {\bf 13}, 17
(73) and J. Low Temp. Phys. {\bf 16}, 87 (74).

\bibitem{thin2}  M. Gabay, Physica (Amsterdam){\bf 235-240C}, 1337 (96); N.
Nagaosa and P.A. Lee Phys. Rev. Lett. {\bf 79}, 3755 (97); G. Khaliullin 
{\it et al.}, Phys. Rev. B {\bf 56}, 11882 (97).

\bibitem{thin}  V. Brunel, M. Bocquet and Th. Jolicoeur, Phys. Rev. Lett. 
{\bf 83}, 2821 (99); I. Affleck and S. Qin condmat/9907284.

\bibitem{allcomm}  H. Alloul {\it et al.}, Phys. Rev. Lett. {\bf 78}, 2494
(97).

\bibitem{isotropic}  We found that $^{7}T_{1}$ is nearly isotropic implying
a nearly isotropic hyperfine coupling together with a nearly isotropic $\chi
^{imp}$. Thus we include the highly overdoped Ca sample with the others.

\bibitem{nmrrev}  A. Rigamonti, F. Borsa and P. Carretta, Rep. Prog. Phys. 
{\bf 61} 1367 (98); C. Berthier {\it et al.}, J. Phys. (Paris) {\bf 12},
2205 (96).

\bibitem{mor}  T. Moriya, J. Phys. Soc. Jpn. {\bf 18}, 516 (63).

\bibitem{sidis}  Y. Sidis {\it et al.}, Phys. Rev. B {\bf 53}, 6811 (96); Y.
Sidis {\it et al.}, Int. J. Mod. Phys. B {\bf 12}, 3330 (98).

\bibitem{epr} The values $\tau$ are also much too short to be related to the narrow EPR
lines observed in LSCO [ A.M. Finkel'stein {\it et al.}, JETP Lett. {\bf 51}, 129
(90) and Physica (Amsterdam) {\bf 168C}, 370 (90)]
which we thus conclude are unrelated to the induced moments
discussed here, and likely originate in a paramagnetic impurity phase.

\bibitem{jan}  A. Janossy, L.C. Brunel and J.R. Cooper, Phys. Rev. B {\bf 54}, 10186 (96).

\bibitem{corr}  We account for the other CuO$_2$ plane of the bilayer using
the approximate corrections\cite{mah2}: $K_{imp} = K_{satellite} -
0.5K_{pure}$ and $1/T_1^{imp} = 1/T_1^{sat} - 0.5/T_1^{pure}$, where the
first correction is much more significant than the second due the large $T_1$
contrast between the outer satellite and the mainline.
\end{thebibliography}
\end{document}